# A Comparison of Nineteen Various Electricity Consumption Forecasting Approaches and Practicing to Five Different Households in Turkey

Benli T.O. [a]

[a] *Graduate School of Clean and Renewable Energies, Hacettepe University, P.O. Box 06800 Beytepe, Ankara, Turkey*

ARTICLE INFO ABSTRACT

The accuracy of the household electricity consumption forecast is vital in taking better cost effective and energy efficient decisions. In order to design accurate, proper and efficient forecasting model, characteristics of the series have to been analyzed. The source of time series data comes from Online Enerjisa System, the system of electrical energy provider in capital of Turkey, which consumers can reach their latest two year period electricity consumptions; in our study the period was May 2014 to May 2016. Various techniques had been applied in order to analyze the data; classical decomposition models; standard typed and also with the centering moving average method, regression equations, exponential smoothing models and ARIMA models. In our study, nine teen different approaches; all of these have at least diversified aspects of methodology, had been compared and the best model for forecasting were decided by considering the smallest values of MAPE, MAD and MSD. As a first step we took the time period May 2014 to May 2016 and found predicted value for June 2016 with the best forecasting model. After finding the best forecasting model and fitted value for June 2016, than validating process had been taken place; we made comparisons to see how well the real value of June 2016 and forecasted value for that specific period matched. Afterwards we made electrical consumption forecast for the following 3 months; June-September 2016 for each of five households individually.

## 1. Introduction

Increase in the world population and requirement of higher life standards cause rapid growth at the amount of electricity usage. Successive planning of energy investment and capacity decisions can help overcoming the sustainability problems. Turkey has limited numbers of local energy sources and she has to import 65% of primary energy to meet the demand [1]. It is expected that Turkey's electrical energy consumption will continue to grow snappily at approximately 8% per year [2]. With the intention of reducing foreign dependency and taking energy efficient decisions, it is obvious that decision takers and policy makers in Turkey are in need of effective forecasting tools and methods which will be developed and executed by researchers. We can say that in the business environment, forecasting is an very important planning tool [3].

Academicians and researchers had put lot of efforts in developing tools and models for forecasting. Analyzing the historical data and apply statistical knowledge in order to relate the predicted values with data in the past and also reflect the characteristics of time series are very crucial. Time series comprised of the data have been collected over time with the continuous period of time. The data which will be analyzed can be in a daily, weekly, monthly or yearly format. Choosing the appropriate period format that will be used in model plays important role on accuracy of models. For the energy efficient concerns, we will provide next 3 months electricity consumption forecasts in order to reveal the potential consumption, cost and savings. Prediction of the electricity consumption of households is important for taking energy efficient decisions. In Turkey the market is structured in according to day ahead pricing system and we see three different periods in one day; namely day, peak and night. It is more effective to make analyze on electrical consumption amounts individually with each period at one day. This kind of approach will help us to understand the characteristics of time series much better and reflect that characterisity to our forecast model in a more accurate way. Temperature variations, people's daily life activities, price of electricity and the people's buying capability, weekend or national holidays, population of households have all impact on the usage of electricity, when analyzing characteristic of usage, those factors have to be considered.

Categorization of electricity consumption forecasting can based on three different time horizons: short term (mainly one day ahead), medium-term (six months to one year) and long-term (one year and more). Several different methods and models have been developed for forecasting, especially for short term periods [4]. Short term electricity consumption forecasting has become very important in today's power industry [5]. Developing an accurate, fast and reliable short term forecasting methodology is important for both the electric utilities and its customers [6]. For the short term forecasts, stationary and non-stationary time series models can be used [7]. Auto-Regressive Integrated Moving Average (ARIMA) is used for forecasting both short-term and long-term periods [8]. ARIMA model is common linear model that have been used in time series forecasting during past three decades [9].

Various methods for statistical forecasting are; regression analysis, classical decomposition method, Box and Jenkins and exponential smoothing techniques. Exponential smoothing techniques are probably the most used method in electricity consumption forecasting. The model's simplicity, computational efficiency and high accuracy lie behind of the popularity of exponential smoothing methods. Being a member of exponential smoothing methods, Holt-Winters exponential smoothing is one of the popular approach for forecasting time series, due to its robust and accurate characterisity [10].

All these different techniques have different characteristics and diversified approach for forecasting the time series data, and give different accuracy. Error measurements are used for determining the accuracy of forecasting models. The technical factors when executing the forecasting model have impact on the accuracy and error measurements of the models. Prediction interval, prediction period, characteristic and size of time series affect the error measurements of various techniques.

In this research for forecasting monochromic, day, peak and night electrical consumptions of five different households, we are interested in nine teen various approaches; which had got different aspects through others, like different seasonality characteristics; twelve and four, different kind of models while applying classical decomposition technique; with additive and multiplicative models, and based on different techniques; standard classical decomposition model, classical decomposition model with centering moving average method, regression equations, single, double and triple exponential (Holt Winters model) smoothing models and lastly ARIMA model. The most suitable forecasting method and the best choice of period were chosen by considering the smallest values of MAPE, MAD and MSD.

The remainder of the paper is structured as follows. Section 2 gives the literature review. Section 3 describes our methodology and the form of the methods we used. Section 4 gave the results of our models, experimental evaluation, also gave the results of validating process and the three month mean fit values of our best forecasting model. In section 5, the discussions part talked about our inferences and section 6 gave the opinion for future works and conclusions.

## 2. Literature Review

Researchers gave lot of effort on studies with the purpose of improving energy consumption forecast accuracy, various models and approaches have been presented. Exemplary, Saab and colleagues [11] studied the forecasting method for monthly electric energy consumption with two different methods; ARIMA and AR(1) in Lebanon. In [12] Zhu, Guo, and Feng used ARIMA and BVAR forecasting methods from the year 1980 to 2009 in China to forecast household energy consumption. Ediger and Akar [13] executed SARIMA (Seasonal ARIMA) methods to predict future fuel energy demand in Turkey from the years 2005 to 2020. For electricity consumption forecasting, D. Srinivasan, C.S. Chang, A.C. Liew [14] presented linear regression models.

Within the purpose of energy consumption forecasting, Bianco V, Manca O, Nardini S. [15] applied gray prediction model. Making short term consumption forecasting in the electricity market of Iran, Zhou P, Ang BW, Poh KL. presented an improved singular spectral analysis method. Forecasting consumption of conventional energy use in India, Kumar and Jain [16] executed Grey-Markov model, Grey-Model with rolling mechanism, and singular spectrum analysis models.

Having the purpose of predicting the electricity consumption in Perlis, Syariza and Norhafiza (2005) compared variouss forecasting methods. However, it is known that Box-Jenkins method is one of the most popular forecasting methods, in their study they found that the Box-Jenkins method is not appropriate to use, also they indicated that regression model is much better for their problem. Afterwards another study came into light, the studies by Taylor (2008) showed that exponential smoothing method is more reliable and appropriate for short term prediction.

## 3. Methodology

### 3.1 Models in the Study

We presented in here different nineteen approaches for forecasting the households' electricity consumption in Ankara in Turkey. Techniques chosen for this study are; standard classical decomposition model (seasonality: 12 or 4), classical decomposition model with centering moving averages method (seasonality: 12 or 4), regression equations methodology (seasonality: 12 or 4), single and double exponential smoothing models (seasonality: 12 or 4) and Holt Winter's method (seasonality: 12 or 4) and ARIMA model methodology (seasonality: 12 or 4).

**Table 1:** Our nine teen various approach for forecasting

| # | Description of forecasting approaches |
|---|---|
| 1 | Classical decomposition with multiplicative model, seasonality: 12 |
| 2 | Classical decomposition with multiplicative model, seasonality: 4 |
| 3 | Classical decomposition with additive model seasonality: 12 |
| 4 | Classical decomposition with additive model seasonality: 4 |
| 5 | Classical decomposition with centering moving averages with multiplicative model, seasonality: 12 |
| 6 | Classical decomposition with centering moving averages multiplicative model, seasonality: 4 |
| 7 | Classical decomposition with centering moving averages with additive model, seasonality: 12 |
| 8 | Classical decomposition with centering moving averages with additive model, seasonality: 4 |
| 9 | Forecasting with regression equation, seasonality: 12 |
| 10 | Forecasting with regression equation, seasonality: 4 |
| 11 | Single exponential smoothing, seasonality: 12 |
| 12 | Single exponential smoothing, seasonality: 4 |
| 13 | Double exponential smoothing with multiplicative model, seasonality: 12 |
| 14 | Double exponential smoothing with additive model, seasonality: 12 |
| 15 | Double exponential smoothing with multiplicative model, seasonality: 4 |
| 16 | Double exponential smoothing with additive model, seasonality: 4 |
| 17 | Forecasting with Holt Winter's model with ideal coefficients, seasonality: 12 |
| 18 | Forecasting with Holt Winter's model with ideal coefficients, seasonality: 4 |
| 19 | ARIMA models |

### 3.2.1 Classical Decomposition Models

In decomposition process, we need to determine the factors that have effect on each value of the time series. Therefore, we identified each component separately in order to show the effect of each component, thus forecasting of future values become possible. Decomposition models are comprised of three components in total; trend, seasonal and irregular components.

Additive components model (observed value stated as $Y_t$) treats the time series values as a sum of three components; seasonality component ($S_t$), trend component ($T_t$) and irregular component ($I_t$). Notation of additive composition model;

$$Y_t = T_t + I_t + S_t \qquad (1)$$

Multiplicative components model (observed value stated as $Y_t$) treats the time series values as the product of the three components; seasonality component ($S_t$), trend component ($T_t$) and irregular component ($I_t$). Notation of multiplicative composition model;

$$Y_t = T_t \times I_t \times S_t \qquad (2)$$

### 3.2.2 Classical Decomposition with Centering Moving Average Model

On the contrary to classical decomposition models, hereby in this model we assigned equal weights to each observation when calculating the seasonal indexes that we will use in forecasting. Notation of centering moving average model, $Y_t$ represents the actual value, $Y_{t+1}$ represents the forecasted value and k represents the number of terms in the moving average, is show below.

$$Y_{t+1} = (Y_t + Y_{t-1} + \ldots + Y_{t-k+1})/k \qquad (3)$$

### 3.3 Regression Equations

In multiple regression models, there are more than one independent variables exist. If it is needed to be clarified that how a dependent variable is related to and independent variable, creating dummy variables will be necessary. Notation of general regression equation form when the seasonality is 4 and 12 is shown below;

$$Y_t = C_0 + T_0 \cdot t + \beta_2.s_2 + \beta_3.s_3 + \beta_4.s_4 \quad \text{(when s = 4)} \qquad (4)$$

$$Y_t = C_0 + T_0 \cdot t + \beta_2.s_2 + \beta_3.s_3 + \beta_4.s_4 + \beta_5.s_5 + \beta_6.s_6 + \beta_7.s_7 + B_8.s_8 + \beta_9.s_9 + \beta_{10}.s_{10} + \beta_{11}.s_{11} + \beta_{12}.s_{12} \quad \text{(when s = 12)} \qquad (5)$$

### 3.4 Exponential Smoothing Models

Exponential smoothing models include; single exponential smoothing, double exponential smoothing, Holt-Winter's smoothing. More recent values have more weight in the exponential smoothing forecasting models. Smoothing of the past values of time series with an exponential decreasing manner is the key for that model.

General notations for the exponential smoothing models are shown below with the smoothing constants; α, β and ɤ. $T_t$ and $C_t$ represented the smoothed trend and constant value individually. $Y_t$ represented real value in that time period and $F_{t+1}$ symbolized forecasted future value and $F_t$ represented forecasted value for the time period of t.

### 3.4.1 Single Exponential Smoothing Model

When there is no linear trend in the time series, single exponential smoothing model will be beneficial. That model is appropriate for short term forecasting.

Forecasting equation with single smoothing constant is shown below.

$$F_{t+1} = \alpha \cdot Y_t + (1-\alpha) \cdot F_t \quad (6)$$

### 3.4.2 Double Exponential Smoothing Model

When increasing or decreasing trend appeared in the time series, modification to the single exponential smoothing model with the intention of adjusting the trend behavior will need to be done. A second smoothing constant, β, is included to account for the trend. Double exponential smoothing model is also appropriate for short term forecasting. Equations for double exponential smoothing model, p letter represents the forecasted period into the future, are shown below.

$$F_{t+1} = C_t + p \cdot T_t \quad (7)$$

### 3.4.3 Holt-Winter's Smoothing Model

When there is both trend and seasonality characteristic appeared in the time series data, double exponential smoothing model can't be used. With the purpose of handling seasonality, we have to add a third parameter, ɤ. Each observation is the product of a non-seasonal value and a seasonal index for that particular period in that technique. $S_t$ represents overall smoothing, $b_t$ represents trend smoothing, $I_t$ represents seasonal smoothing, L represents the length of periods and m represents the number of period that will be used in forecasting.

$$F_{t+m} = (S_t + m \cdot b_t) \cdot I_{t-L+m} \quad (8)$$

## 3.5 ARIMA Models

As well as the exponential smoothing models are accepted as and effective in short term forecasting, ARIMA models can also be used in short term forecasting and can generate good accuracies. ARIMA models use iterative approach in identifying the best possible model that will give the smallest error measurements.

### 3.5.1 Autoregressive Models

Regression model with lagged values of the dependent variable can be presented as independent variable in the autoregressive models, and that models are suitable for to be used in stationary time series. A first order autoregressive model equation is shown below, where Φ represents the coefficients to be estimated and $Y_t$ and lagged ones represent the response variable at time t.

$$Y_t = \Phi_0 + \Phi_1 \cdot Y_{t-1} + \Phi_2 \cdot Y_{t-2} + \ldots + \Phi_p \cdot Y_{t-p} \quad (9)$$

### 3.5.2 Moving Average Models

The deviation between response of the model and mean of the model is linear combination of current and past errors. We referred to that situation as moving average. With the coefficients of: ω, the constant mean of the process: μ, the error term in that current time and also past errors, $Y_t$, the response variable at time t and q, as the number of past error terms, the general notation of the moving average model is shown below.

$$Y_t = \mu + \varepsilon_t - \omega_1 \cdot \varepsilon_{t-1} - \omega_2 \cdot \varepsilon_{t-2} - \ldots - \omega_q \cdot \varepsilon_{t-q} \quad (10)$$

### 3.5.3 Autoregressive Moving Average Models

When the moving average terms and autoregressive terms combine, we will have autoregressive moving average model with the p and q order, gives the order of autoregressive and moving average part correspondingly. The general equation of autoregressive moving average model is shown below.

$$Y_t = \Phi_0 + \Phi_1 \cdot Y_{t-1} + \Phi_2 \cdot Y_{t-2} + \ldots + \Phi_p \cdot Y_{t-p} + \varepsilon_t - \omega_1 \cdot \varepsilon_{t-1} - \omega_2 \cdot \varepsilon_{t-2} - \ldots - \omega_q \cdot \varepsilon_{t-q} \quad (11)$$

## 3.6. Measuring Forecast Error

### 3.6.1 Mean Absolute Deviation (MAD)

Mean Absolute Deviation (MAD) calculates the forecast accuracy by averaging the absolute values of the forecast errors.

$$MAD = \frac{1}{n} \sum_{t=1}^{n} |Y_t - \hat{Y}_t| \quad (12)$$

### 3.6.2 Mean Squared Deviation (MSD)

Because of the squared term exist in the equation, we can say that this method penalizes large forecasting errors.

$$MSD = \frac{1}{n} \sum_{t=1}^{n} (Y_t - \hat{Y}_t)^2 \quad (13)$$

### 3.6.3 Mean Absolute Percentage Error (MAPE)

There can be some cases occur that it will be more beneficial to calculate the forecast errors in terms of percentages rather than amounts. After calculating the absolute error in each period, dividing this by actual value for that period, we hereby calculated average of absolute percentage errors. Then, the result at the final will be multiplied by hundred, and that will allow us to express the error as percentage, MAPE.

$$MAPE = \frac{1}{n} \sum_{t=1}^{n} \frac{|Y_t - \hat{Y}_t|}{|Y_t|} \quad (15)$$

## 3.7 Methodology Description

The population of this research is electrical consumption data available in the period from May 2014 to May 2016 in total account for five households in Ankara, Turkey. Three households locate in the Birlik district which is in Cankaya and the others; one is under Kirkkonaklar, again in Cankaya and Eryaman district in Yenimahalle district individually. The electrical consumption data of houses collected from Enerjisa company's website called "Online Enerjisa", due to fact that Enerjisa Company was endowed with authority of providing electrical energy to consumers in Ankara by the government.

In total nine teen approaches for forecasting the electrical consumption data were executed for day, peak, night and total time periods for all five houses. We found error measures for all nineteen approaches and afterwards comparing the MAPE, MAD and MSD values, we decided which approaches were the best for forecasting the electrical consumption data for each house. After finding the best appropriate models for forecasting for five house individually, we compared our forecasted value of June 2016 with real electricity consumption data in all five houses, in that step we made our validating process of our best appropriate model. That step revealed the success criterion of out method we had chosen, method will smaller error measures, by letting us to see whether the forecasted June 2016 value were in what extend correct or not. After bringing the success of our method to light, we made electricity consumption forecast for 3 months ahead; July, August and September 2016.

## 4. Experimental Evaluation

We conducted nineteen different forecasting approaches for five house; four different electricity time period for each; total, day, peak and night. You can choose your electricity pricing system whether you can choose monochromic time period, namely as for total electricity and the time period of total electricity: 00:00 am- 24:00 pm, or you can choose multi time period pricing system namely; day, peak and night and time intervals for each corresponds to; day: 06:00 am- 17:00 pm, peak: 17:00 pm – 22:00 pm, night: 22:00 pm – 06:00 pm. All of those different time periods are priced differently, if you choose monochromic time period tariff you will pay different, if you will choose multi time period electricity tariff you will pay different for the total electricity consumption you make. Benli and Sengul executed extensive study about one year ahead electricity tariff price forecasting in Turkey. In that study, we were merely interested in with the forecasting of the electricity consumption not the price. We analyzed the electricity time series data of five houses with the time period of May 2014 – May 2016, fact that consumers can only reach last two year electrical consumption from the supplier's system, Enerjisa company, limited us doing analysis and forecasting with more data.

Our three houses locate in Birlik district in Cankaya and one of them is in Kirkkonaklar district in Cankaya and the other one locates in Eryaman district in Yenimahalle. House 1, house 3 and house 4 is a penthouse apartment, house 2 is also penthouse apartment but it's a working office and house 5 is three floored villa. We analyzed the electrical consumption data series for total, day, peak and night period time for each of the five households with the time period of May 2014 to May 2016. Time series data of total, day, peak and night time electrical consumption for two years period for all of the five households can be seen in the supplementary document. We saw stationary behavior in all of the five houses' electrical consumption time series when we looked into big picture in the graphs. Constant values for the mean of the electrical consumption for all of the five houses are; ≈ 460 kWh, ≈ 350 kWh, ≈ 260 kWh, ≈ 242 kWh and ≈ 370 kWh; house 1, house 2, house 3, house 4 and house 5, individually. The total amount of two year electrical consumption of five houses are; 11311.01 kWh, 10081.593 kWh, 6525.989 kWh, 6087.457 kWh, 9787.688 kWh; house 1, house 2, house 3, house 4 and house 5, individually.

When we wanted to analyze the consistency of electrical consumption time series data of five households, we saw that house 3 and house 4's consumption behaviors looked so similar, they both showed stability in their characteristic and we didn't see unexpected peaks, on the final house 3's total electricity consumption was 438.532 kWh higher than house 4.

House 1 and house 5's behavior looked like to each other. House 1's total consumption was significantly higher until to January 2015, after that month they followed closely except the period of November 2015 to February 2016. In that period, house 5's consumption was higher due to running the heater for the dog in the garden. Among these all of the five households, house 2 showed unstable fluctuations. The result was expected because the population in the house clearly higher than the others, population based consumption differed much more than from the other households. Like house 5, in the period of November 2015 to March 2016, when the temperature went down in Turkey, due to the office population lacked obtaining the necessity heating level, they used many heaters in the office environment. That was the reason of we saw huge peak in the electricity consumption data of the house 2. All of the electrical consumption time series data for time period of total, day, peak and night, can be seen in supplementary document. Comparison of all of the five different households can be seen in the Figure 1.

**Table 2:** Error measures of best models with the seasonality of 12.

|  |  | Total | Day | Peak | Night |
|---|---|---|---|---|---|
| House 1 Best model # : 19 | MAPE | 10,6755 | 10,6127 | 10,7353 | 8,2974 |
|  | MAD | 48,6213 | 22,0032 | 12,5734 | 10,2566 |
|  | MSD | 5641,423 | 1286,81 | 441,221 | 366,999 |
| House 2 Total, peak, night : # 9 Day : # 5 | MAPE | 19,4300 | 22,1978 | 16,8519 | 12,6883 |
|  | MAD | 73,2833 | 55,1862 | 12,8786 | 7,8425 |
|  | MSD | 6567,046 | 5996,29 | 188,919 | 83,7806 |
| House 3 Day, peak, night : # 9 Total : # 1 | MAPE | 4,1437 | 3,9830 | 5,9228 | 4,4689 |
|  | MAD | 10,8162 | 4,6864 | 4,3292 | 3,2874 |
|  | MSD | 256,8618 | 31,6378 | 26,7224 | 17,7851 |
| House 4 Best model # :1 | MAPE | 6,9406 |  |  |  |
|  | MAD | 18,6964 |  |  |  |
|  | MSD | 1300,325 |  |  |  |
| House 5 Best model # : 19 | MAPE | 8,826 |  |  |  |
|  | MAD | 40,2665 |  |  |  |
|  | MSD | 5930,820 |  |  |  |

We implemented our nineteen approaches forecasting methods for the electrical consumption data of; total, day, peak and night for house 1, house 2 and house 3. We executed our nineteen various methods for the electrical total consumption of house 4 and house 5, because these households didn't have electronic counter for last two years, they had for only last couple of months, so we couldn't get the consumptions for the separate time periods and thus we weren't able to execute our forecasting methods due to lacking of time series data.

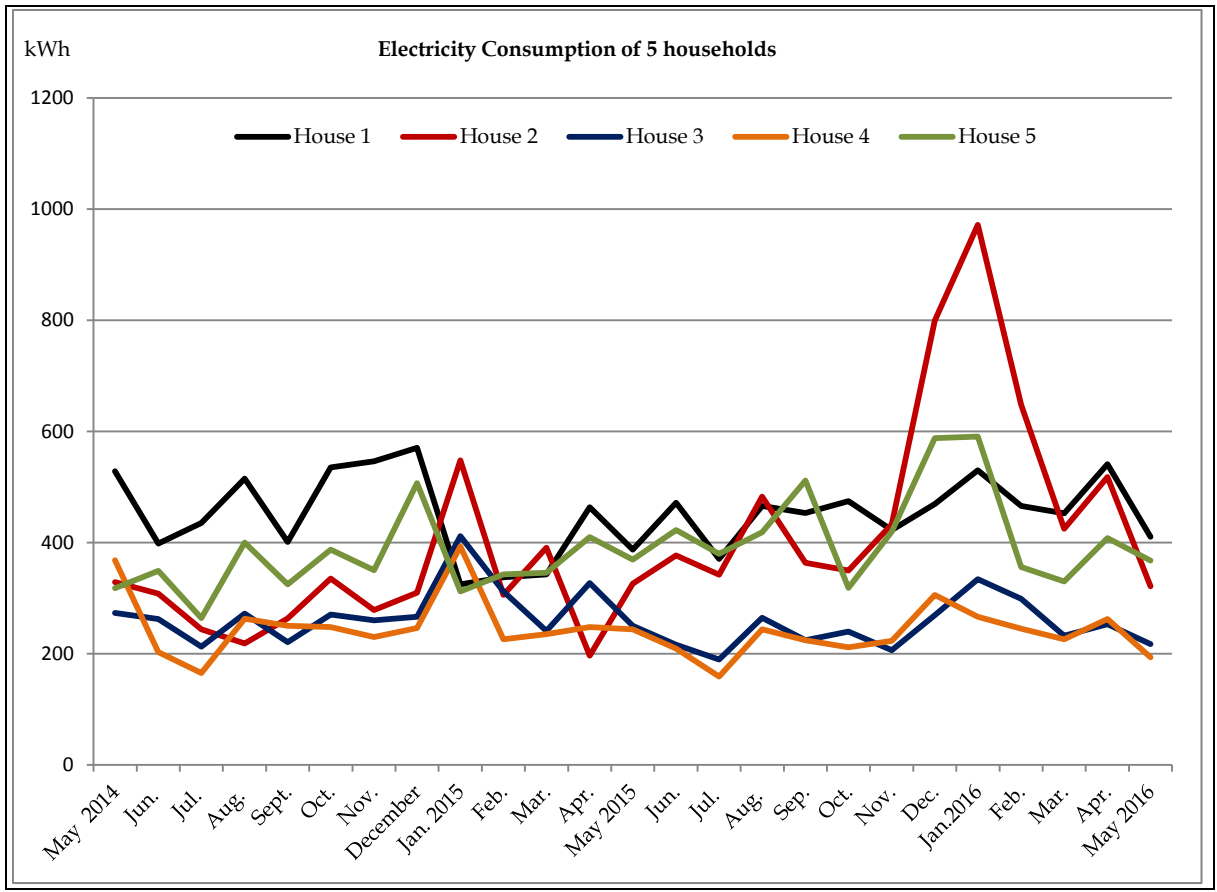

Fig. 1: Total Electricity Consumption Time Series of 5 Households

Regardless, we executed all different nineteen approaches for forecasting, with the intention of validating our forecasted values with real values, we hereby compared the models that have seasonality of 12. Due to we had only have June values, we can only compared the June 2016 real value with forecasted one you can see the results in next section of this study.

Error measures with the Best forecasting method for the house 1 that we found after the execution with our approaches, for the time periods of total, day, night and peak is our approach number 19; ARIMA model with the seasonality of twelve. ARIMA (0, 0, 0) (1, 0, 0) $_{12}$ model was chosen to apply for the forecasting for house 1.

Best forecasting method for the house 2 after the implementing our nineteen various approaches which have the seasonality of twelve; regression equation forecasting approach was the best choice for the time periods of total, peak and night, and the classical decomposition with centering moving average with multiplicative approach model was the best choice for the time period for day.

Best forecasting method for the house 3 with the seasonality of twelve for the time periods of day, peak and night was our approach number 9, regression equation forecasting approach, however best forecasting method for the time period of total was our approach number 1; classical decomposition with multiplicative model. As we said before, due to not using electronic counter on house 4, we only executed our models for the time period of total. The best appropriate model which has the seasonality of 12 was our approach number 1; classical decomposition model with multiplicative approach.

Like house 4, we only implemented our approaches only for the time period of total in house 5, so the best appropriate model that has the seasonality of 12 was our approach number 19; ARIMA model with the seasonality of twelve. ARIMA (0, 0, 0) (1, 0, 0) $_{12}$ model was chosen to apply for the forecasting for house 5.

MAPE, MAD and MSD error measures of the best appropriate models that have the seasonality of 12 and used for forecasting for the time periods of total, day, peak and night were shown in Table 2. You can also reach all nineteen approaches error measures for the time periods of total, day, peak and night for the houses 1-3 and for the time period of total for the house 4 and house 5 in the supplementary document. In Table 4, we also shared the best forecasting approaches together with the seasonality 4 and 12, we compared all of them between each other in order to find the best model which had the smallest error measures and we used that model for forecasting the 3 months future values.

## 4.1 Validation of Our Best Forecasting Models

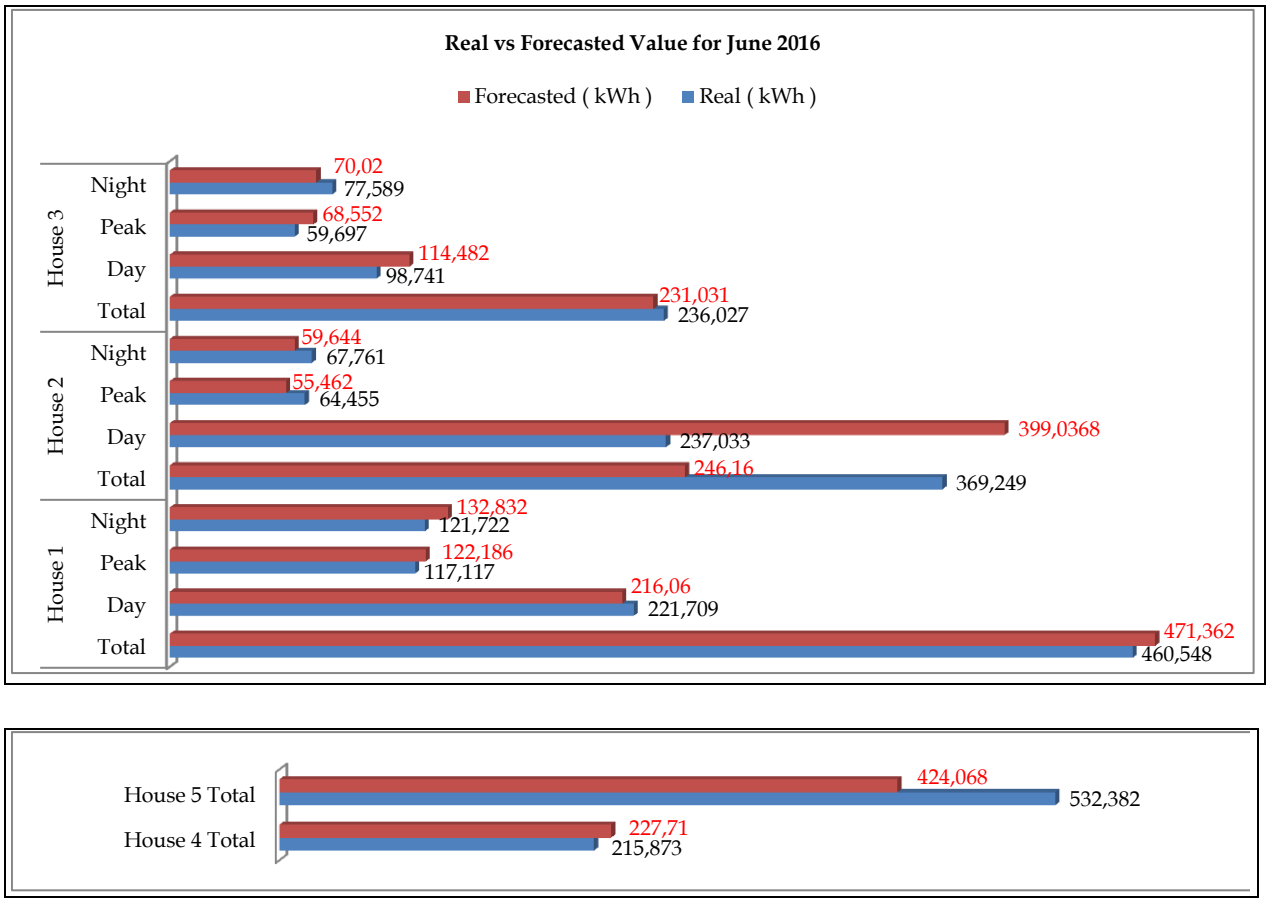

Fig. 2: Real vs Forecasted Values Comparison for the June 2016 for all five households

We compared how well the forecasted 2016 June value fit with the real value; we did comparison for all of the five houses for all of time period we analyzed. Approximation percentage errors were shown in Table 3. As we expressed before, in house 2 (office) and also house 5 fluctuations were significantly higher than the other households. It was difficult to predict the electrical consumption value for those houses than house 1, house 3 and house 4. Approximation percentage errors of house 1 were the best with the smallest measures among through others. The worst approximation percentage errors were belonged to house 2 (office), following that to house 5 for the June 2016. Before comparing the error measures, we accepted that the model with the smaller error measures should gave more accurate fit values. House 1's approximation percentage errors were; 2.3480%, - 2.5479%, 4.3281%, 9.1273%, for the time periods of; total, day, peak, night, correspondingly. House 2's approximation percentage errors were; -33.3349%, 68.3465%, -13.9523%, -11.9788%, for the time periods of; total, day, peak, night, correspondingly. House 1's MAPE values were 10.6755, 10.6127, 10.7353, 8.2974, for the time periods of; total, day, peak, night, correspondingly. House 2's MAPE values were 19.430, 22.1978, 16.8519, 12.6883, for the time periods of; total, day, peak, night, correspondingly.

**Table 3:** Approximation percentage errors for the June 2016 for all of the five houses.

|  | Total (%) | Day (%) | Peak (%) | Night (%) |
|---|---|---|---|---|
| House 1 | 2, 3480 | - 2, 5479 | 4, 3281 | 9, 1273 |
| House 2 (Office) | - 33, 3349 | 68, 3465 | - 13, 9523 | - 11, 9788 |
| House 3 | - 2, 1167 | 15, 9417 | 14, 8332 | - 9, 7552 |
| House 4 | 5, 4833 | | | |
| House 5 | - 20, 3451 | | | |

In Figure 2, real versus forecasted values comparison for the June 2016 were shown. Except day and total time periods for house 2 and total time period for house 5, the other houses' and periods' fit values were in the acceptable range, but still need to improvement obviously.

### 4.2 Future Forecast Results

As we stated before, we implemented our nineteen approaches forecasting methods for the electrical consumption data of; total, day, peak and night for house 1, house 2 and house 3, and we executed our nineteen various methods for the electrical total consumption of house 4 and house 5.

Our models whose seasonality was 12 used to predict twelve separate months' forecasts. We compared the best forecasting models whose seasonality was 12 because we had only got June 2016 electrical consumption amount to validate the forecasted June 2016 result with our best approach. After validation step, that time we compared all of our nineteen approaches for forecasting the electrical consumptions of five different households with the related time periods. Including the seasonality of four, and comparing the all various nineteen approaches' forecasting error results, best forecasting methods' error results for all of the houses for the related time period were shown in Table 4. All different nineteen approaches' forecasting results can be seen in our supplementary document.

All the best forecasting methods for the related time periods for the five households had got the seasonality behavior of 4. That situation fathomed out that if the time period of data we will use in analyzing process and forecasting is short, it will be better to use forecasting approaches which had got the seasonality of four and produce time series data on the quarterly basis. In this study our models produced three monthly average data of electrical consumption in all five households in the related period of time. The three months', July, August and September 2016, mean fit values for all of the five households with the related time periods were shown in Table 5.

### 5. Discussions

It may be effective that sometimes people test various different approaches for solving the problem, comparing the results and try to create new attitude. We hereby test nineteen various approaches in order to find which one can be effective tool to forecast the five different households, one of them is office, with the two different seasonality; 4 and 12.

There are various forecasting techniques, the models we used in this study were: classical decomposition model with centering moving averages method, regression equations methodology, single and double exponential smoothing models and Holt Winter's method and ARIMA model methodology.

Minimal data requirements for; ARIMA seasonal models, classical decomposition models, seasonal exponential smoothing models, moving average models are; 3, 5, 2, 4 correspondingly. As we stated before; ARIMA model can be used for short term forecasting, exponential smoothing techniques are common in electricity consumption forecasting, regression models can also be used in electrical consumption forecasting. We could only use two year data due to that electricity provider only provide two year electrical consumptions in the online system.

Firstly, we compared the error results of forecasting methods which had got seasonality of 12 in order to validate our best model's fit value with real June 2016 value. When the seasonality was 12, the best model was ARIMA model for the house 1 for all the time periods. Model # 9 was the best forecasting model for house 2 for the periods of total, day and night and model # 5 was the best forecasting model for the peak period. For the house 3, model # 9 was the best method for the time periods of day, peak and night and model # 1 was the best forecast model for the total time period. For the house 4, model # 1 was the best total time period forecasting approach and for the house 5, model # 19 was the best forecasting approach for the total time period.

When we computed the validation step and compared the approximation percentage errors, approximation percentage errors of house 1 were the best with the smallest measures among through others. The worst approximation percentage errors were belonged to house 2 (office), following that to house 5 for the June 2016.

In order to start forecasting step, that time we compared all various nineteen approaches which had got seasonality of 4 and 12. Among these, model # 10 was the best model that can be used for forecasting the three month mean electrical consumptions of house 3 for all the time period, for house 4 and house 5 with the time period of total, for house 2 with the time period of night, for house 1 with the time periods of total, peak and night. Model # 18 was the best approach for forecasting the mean electrical consumption of house 2 for the time periods of total and day, and for house 1 for the time period of day. Lastly, model # 6 was the best model for forecasting the peak time period for house 2.

### 6. Conclusions

Having smaller error measures gives the sign of our forecasting model can be effective tool for predicting the future values. Analyzing the time series data help us to understand the consumers' electrical consumption behavior. Big fluctuations in the time series data is prohibiting factor that constrain us from getting effective forecasting tool, in that time we will need to use professional forecasting methods with inputs. In our study, we can only validate our best model working performance with June 2016 results. Despite having only two year electrical consumption series data, we can say that our approaches to creating effective forecasting tool show promise. Benli and Sengul showed that increasing the time series period obviously help us with getting better forecast results, in that concept we need to complete this study again with more time series data for different houses and need to compare the effect of the length of time series period.

**Table 4:** Best forecasting method's error measures among all the nineteen various forecasting approaches

| | | MAPE | MAD | MSD |
|---|---|---|---|---|
| House 1 | Total best model # : 10 | 5,50401 | 24,11681 | 821,34792 |
| House 1 | Day best model # : 18 | 4,90632 | 10,28622 | 270,13439 |
| House 1 | Peak best model # : 10 | 5,60612 | 6,2885 | 46,83094 |
| House 1 | Night best model # : 10 | 3,2397 | 3,99837 | 23,60733 |
| House 2 | Total best model # : 18 | 11,39577 | 44,29804 | 3559,34251 |
| House 2 | Day best model # : 18 | 13,61238 | 32,43054 | 1806,3243 |
| House 2 | Peak best model # :6 | 6,49979 | 4,48154 | 29,3277 |
| House 2 | Night best model # : 10 | 8,14843 | 5,32491 | 33,74048 |
| House 3 | Total best model # : 10 | 1,67005 | 4,40827 | 31,29829 |
| House 3 | Day best model # : 10 | 2,47879 | 2,9462 | 9,94511 |
| House 3 | Peak best model # : 10 | 1,45367 | 1,08145 | 1,35084 |
| House 3 | Night best model # : 10 | 3,98381 | 2,9322 | 11,03461 |
| House 4 | Total best model # : 10 | 3,11374 | 7,3356 | 81,89 |
| House 5 | Total best model # : 10 | 5,60454 | 22,35366 | 692,21585 |

**Table 5:** Three month mean fit values (electricity consumption) for all of the five households with the related time periods

|  |  | July - August 2016 ( kWh ) |
|---|---|---|
| House 1 | Total | 444.81 |
| House 1 | Day | 194.166 |
| House 1 | Peak | 113 |
| House 1 | Night | 119.89 |
| House 2 | Total | 572.981 |
| House 2 | Day | 370.702 |
| House 2 | Peak | 116.344 |
| House 2 | Night | 94.08 |
| House 3 | Total | 190.24 |
| House 3 | Day | 89.891 |
| House 3 | Peak | 57.511 |
| House 3 | Night | 45.75 |
| House 4 | Total | 192.63 |
| House 5 | Total | 450.33 |

# REFERENCES

[1] Oulata, T.; 2003; Energy sector and wind energy potential in Turkey; **Renewable and Sustainable Energy Reviews**; Vol: 7; No: 6; pp: 469-484. DOI=10.1016/S1364-0321(03)00090-X.

[2] Kucukdeniz, T.; 2010; Long term electricity demand forecasting: an alternative approach with support vector machines; **Istanbul University of Engineering Sciences**; Vol: 1; pp: 45-53.

[3] Bianco V, Manca O, Nardini S. Electricity consumption forecasting in Italy using linear regression models. Energy 2009;34(9):1413e21.

[4] Taylor, G., Irving, M. and Hu, L.L.; 2008, A Fuzzy logic based bidding strategy for participants in the UK electricity market; Padova, 1-5.

[5] W. Charytoniuk, and M.S. Chen, "Very Short-Term Load Forecasting Using Artificial Neural Networks", IEEE Transactions on Power Sys¬tems, vol. 15, no. 1, pp. 263-268, 2000.

[6] Patell, D.P, Tiwari. A., Dubey, V.; 2013; An Analysis of Short Term Load Forecasting by Using Time Series Analysis; International Journal of Research in Computer and Communication Technology, Vol: 2; No: 2; pp: 48-53.

[7] Zareipour, H., Janjani, A., Leung, H., Motamedi, A., Schellenberg, A.; 2011;  Classification of future electricity market prices;  IEEE Trans. Power Syst**,** Vol: 26, pp: 165-173.

[8] Niu, J., Xu, Z., Zhao, J., Shao, Z., and Qian, J.; 2010; Model predictive control with an on-line identification model of a supply chain unit.

[9] Makridakis, S., S.C. Wheelwright and R.J. Hyndman; 1998; Forecasting methods and applications; 3rd. ed. Wiley, Inc., New York.

[10] Kotillovâ, A.; 2011; Very Short-Term Load Forecasting Using Exponential Smoothing and ARIMA Models; Journal of Information, Control and Management Systems, Vol: 9, No: 2; pp: 85-92.

[11] Taylor, J. W.; 2003, Short-Term Electricity Demand Forecasting Using Double Seasonal Exponential Smoothing; The Journal of the Operational Research Society, Vol: 54, No: 8, pp: 799-805.

[12] Samer Saab, Elie Badr and George Nasr, "Univariate modeling and forecasting of energy consumption: the case of electricity in Lebanon," Energy, vol.26, 2001, pp. 1-14.

[13] Qing Zhu, Yujing Guo, Genfu Feng, "Household energy consumption in China forecasting with BVAR model up to 2015,"  2012 Fifth International Joint Conference on Computational Sciences and Optimization, 2012.

[14] Volkan Ş. Ediger, Sertaç Akar, "ARIMA forecasting of primary energy demand by fuel in Turkey," Energy Policy, vol.35, 2007, pp. 1701-1708.

[15] D. Srinivasan, C.S. Chang, A.C. Liew, Survey of hybrid fuzzy neural approaches to electrical load forecasting, in: Proceedings on IEEE International Conference on Systems, Man and Cybernetics, Part 5 Vancouver, BC, 1995, pp.4004-4008.

[16] Zhou P, Ang BW, Poh KL. A trigonometric grey prediction approach to forecastingelectricity demand. Energy 2006;31(14):2839e47.

[17] Kumar U, Jain VK. Time series models to forecast energy consumption in India. Energy 2010;35(4):1709e16.